\documentclass[12pt]{article}
\usepackage{epsfig,psfrag,amsmath,amssymb}
\usepackage{axodraw,color}

\newcommand{\rd}{\mathrm{d}}
\newcommand{\alp}{\bar{\alpha}}
\newcommand{\Df}{D_\mathrm{fin}}
\newcommand{\lqcd}{\Lambda_\mathrm{QCD}}

\newcommand{\hhs}{s}
\newcommand{\hht}{t}
\newcommand{\hhu}{u}
\newcommand{\JC}{J_{\mathrm C} }

\begin{document}

\begin{titlepage}

\begin{flushright}
\end{flushright}

\vspace{-0.3cm}
\begin{center}
\Large\bf
Resummation and NLO Matching of Event Shapes \\ 
with Effective Field Theory
\end{center}

\vspace{0.4cm}
\begin{center}
{\sc Matthew D. Schwartz}\\
\vspace{0.4cm}
{Johns Hopkins University \\
 Baltimore, MD, USA\\[0.3cm]}
\end{center}

\vspace{0.2cm}
\begin{abstract}
\vspace{0.2cm}
\noindent 
The resummed differential thrust rate in $e^+ e^-$ annihilation is calculated
using Soft-Collinear Effective Theory (SCET). 
The resulting distribution in the
two-jet region ($T \sim 1)$ is found to agree
with the corresponding expression derived by the standard approach.
A matching procedure to account for finite corrections
at $T < 1$ is then described. There are two important
advantages of the SCET approach. 
First, SCET manifests a
dynamical seesaw scale $q = p^2/Q$ in addition to the center-of-mass
energy $Q$ and the jet mass scale $p \sim Q \sqrt{(1 - T)}$. Thus, the
resummation of logs of $p / q$ can be cleanly distinguished from the resummation
of logs of $Q / p$. Second, finite parts of loop amplitudes appear in specific
places in the perturbative distribution: in the matching to the hard function,
at the scale $Q$, in matching to the jet function, at the scale $p$, and in
matching to the soft function, at the scale $q$. This allows for a consistent
merger of fixed order corrections and resummation. In particular, the total
NLO $e^+ e^-$ cross section is reproduced from these finite parts without
having to perform additional infrared regulation.
\end{abstract}
\vfil

\end{titlepage}

\section{Introduction}

Quantum Chromodynamics is a perturbative field theory for $\alpha_s < 1$,
corresponding to energies above $\lqcd \sim 500$ MeV. However,
setting up a good perturbation expansion is more difficult than simply working
order by order in $\alpha_s$. The difficulty is that when computing a quantity
with more than one scale, logarithms of ratios of those scales appear which
invalidate the naive perturbation expansion. For example, in $e^+ e^-$
collisions at center-of-mass energy $Q$, we might look for the distribution of
jets as a function of the invariant mass $p^2$ of the jet. Then the
differential cross section will have a correction of the form $\alpha_s \log^2
\frac{p^2}{Q^2}$. Even for $\alpha_s < 1$ these large logarithms can dominate
if $p^2$ is sufficiently small. Luckily, these logarithms appear only in
certain combinations even at higher order, so that all the terms of the form
$(\alpha_s \log^2 \frac{p^2}{Q^2})^n$ can be (re)summed at once. However,
understanding which terms will appear, how to resum them, and how to combine
the resummed result with fixed order results can be quite complicated. It is
the goal of this paper to show how it can be done using effective field theory
techniques.

In this paper, resummation and matching of event shapes are studied using
Soft-Collinear Effective Theory 
(SCET)~\cite{Bauer:2000ew,Bauer:2000yr,Bauer:2001yt,Beneke:2002ph,Bauer:2002nz}.
Event shapes are observables which are
sensitive to the overall distribution of final state particles, and therefore
involve both short and long distance physics. We will consider mainly the
event shape $\tau = 1 - T$ where the thrust $T$ is defined by
\begin{equation}
  T = \max_{\mathbf{n}} \frac{\sum_i | \mathbf{p}_i \cdot \mathbf{n}
  |}{\sum_i | \mathbf{p}_i |}
\end{equation}
summing over all momentum $3$-vectors $\mathbf{p}_i$ in the event and
maximizing over unit 3-vectors $\mathbf{n}$. In the threshold region near
$\tau = 0$ large logarithms of the form $\alpha_s \log^2 \tau$ appear at fixed
order in perturbation theory. The resummed result will be valid even if
$\alpha_s \log^2 \tau$ is large, as long as $\alpha_s$ and $\alpha_s \log
\tau$ are small.

Effective field theories
provide a systematic approach to resummation.
They separate out physics at a hard underlying scale $Q$ from physics
associated with a scale of interest $p^2 \sim Q^2 \tau$ and from even lower
scales. At each scale a separate matching calculation is done which
is independent of physics at asymptotically lower or higher energy. Then the scale dependence is
calculated using the renormalization group. In this way, large logarithms
cannot appear because no two largely separated scales are accessible to the
theory at the same time. The Soft-Collinear Effective Theory works by
separating the degrees of freedom of QCD into soft modes and collinear modes
in different directions. The relevant scales are then
associated with the observable of interest,
such as $Q^2 \tau$, or with the large components of the collinear modes.

The resummation of event shapes in $e^+ e^-$ is not currently of extreme importance
phenomenologically. It nevertheless provides a clean arena (as compared to
hadron collisions) to explore resummation and matching. The SCET techniques
and most of the formulas we discuss here were originally developed for
$B$-physics, such as resummation in $b \rightarrow s \gamma$
decays~\cite{Bauer:2000ew,Bauer:2003pi,Bosch:2004th,Neubert:2004dd}.
They have also been applied to the study of deep-inelastic-scattering
near $x = 1$~\cite{Manohar:2003vb,Becher:2006nr,Becher:2006mr}
and to the production of massive jets initiated by top quark
decays~\cite{Fleming:2007qr}.
One convenient feature of effective field theories is factorization,
which allows us to use objects, such as soft and jet functions, calculated in
one process to study another. Thus, most of the hard work required for
calculation of the distributions we describe here can be extracted from the
literature. Nevertheless, there are certain features, in particular the NLO
matching step, for which event shapes are uniquely illuminating.

The breakdown of naive perturbation theory due to the appearance of large
logarithms is independent of $\alpha_s$ blowing up and of non-perturbative
effects. To emphasize this point, and to simplify the resummed expressions,
the running of $\alpha_s$ will be simply turned off (setting $\beta = 0$) for most
of this paper. Working in the conformal limit removes one scale
($\lqcd$) from the problem and thus clarifies which large logs
are being resummed. It is not hard to turn $\beta$ back on, as will be
shown in Section~\ref{sec:gen}. Also, we will be including all one-loop
results but no two-loop results. Thus, our expressions will not
contain a complete next-to-leading log resummation, which should 
also resum the two-loop double logs.

There are three steps involved in the calculation of event shapes in SCET,
associated with three separate scales. At the hard scale $\mu_h \sim Q$, QCD is
matched onto SCET by demanding parton level matrix elements in the two theories be the
same. This can be done at leading order by matching onto an operator
$\mathcal{O}_2$ with two collinear fields, or at next-to-leading order by
matching in addition to an operator $\mathcal{O}_3$ with collinear fields in
three directions~\cite{Bauer:2006mk,Bauer:2006qp}. 
The threshold resummation only involves $\mathcal{O}_2$ as
the matrix elements of $\mathcal{O}_3$ vanish near $\tau =
0$. (An alternative approach to ${\mathcal O}_3$ matching is described 
in~\cite{Trott:2006bk}).
At the scale $\mu_j \sim Q \sqrt{\tau}$, the collinear fields freeze and
can be removed from the theory by integrating them out. This results in the
a jet function $J (p^2)$. Finally, even though we are interested
in a distribution at the scale $Q\sqrt{\tau}$, 
the soft degrees of freedom remain
relevant down to a scale $\mu_s = Q \tau$, after which they too can be
integrated out of the theory. The fixed order result will have large
logarithms of $\mu_h / \mu_j$ and $\mu_j / \mu_s$, but in the resummed result
all these logs are exponentiated and innocuous.

Before we present the factorization formula and calculate the thrust
distribution in SCET, we will review the way resummation of thrust 
is traditionally handled.
There are a number of ways to resum event 
shapes~\cite{Catani:1992ua, 
Catani:1991kz, 
Catani:1996yz, 
Dokshitzer:1998kz, 
Kidonakis:1998bk, 
Gardi:2001ny, 
Berger:2003iw}.
Since the focus of this work is on
comparisons to SCET, we will confine our attention to the original approach of
Catani, Trentadue, Turnock and Webber~\cite{Catani:1992ua}, which 
will be referred to as CTTW throughout. 
Moreover, most of the
other approaches reduce to~\cite{Catani:1992ua} 
at next-to-leading fixed order in $\alpha_s$ (NLO) and to leading log,
so no significant
loss of generality is sustained. We will find that in the two-jet limit SCET
also agrees with CTTW to the order we are working, although the resummed
expressions are not exactly the same.

A more significant difference is in the NLO matching. A critical advantage of
the effective theory approach is that the finite parts of loop amplitudes are
automatically incorporated into the perturbative expressions. For example, the
total NLO cross section for $e^+ e^-$ is reproduced by combining the finite
parts of the hard, jet, and soft functions and a
contribution from a finite integral over higher-order operators in SCET.
This does not necessarily entail less work than in calculating the total cross
section through the traditional combination of real and virtual contributions.
However, due to factorization,
the infrared divergent contributions which are absorbed into the jet and soft
functions are universal and thus the could potentially be used for many
processes.

\section{Perturbative QCD}

In this section, we will review some basic results from QCD on thrust, and the
resummed expressions presented in~\cite{Catani:1992ua}.

To begin, consider the parton model description of $e^+ e^-$ annihilation. At
order $\alpha_s^0$, the only process which contributes is $e^+ e^- \rightarrow
\bar{q} q$. These two jets have no structure and hence the cross section is
simply $d \sigma / d \tau = \sigma_0 \delta (\tau)$, where $\sigma_0$ is 
leading order total $e^+e^-$ annihilation cross section.

At order
$\alpha_s^1$, there are two $e^+ e^- \rightarrow \bar{q} q g$ diagrams which
contribute
\begin{equation}
d\sigma_\mathrm{parton} \sim
\left |
\SetScale{0.5}
\fcolorbox{white}{white}{
  \begin{picture}(50,25) (0,3)
    \SetWidth{0.5}
    \SetColor{Black}
    \LongArrow(45,30)(0,60)
    \Photon(45,30)(45,-15){4}{3}
    \LongArrow(45,30)(90,60)
    \Gluon(72,48)(45,75){4.5}{2.17}
    \Gluon(72,48)(45,75){4.5}{2.17}
  \end{picture}
}
+
\fcolorbox{white}{white}{
  \begin{picture}(50,25) (0,3)
    \SetWidth{0.5}
    \SetColor{Black}
    \LongArrow(45,31)(0,61)
    \Photon(45,31)(45,-14){4}{3}
    \LongArrow(45,31)(90,61)
    \Gluon(21,48)(45,76){4.5}{2.05}
  \end{picture}
}
\right|^2 ,
\end{equation}
where the photon line on the bottom is the $e^+ \gamma^\mu e^-$ current coming in. The differential cross section is
\begin{equation}
\left[  \frac{1}{\sigma_0} 
\frac{\rd^2 \sigma}{\rd \hhs\, \rd \hht}
\right]_{\mathrm{parton}} = \delta (\hhs)
  \delta (t) + \alp \frac{\hhs^2 + \hht^2 + 2 \hhu}{\hhs\,\hht}
 \label{sigparton}
\end{equation}
where we have defined
\begin{equation}
  \alp \equiv \frac{2 \alpha_s}{3 \pi}
\end{equation}
and the reduced Mandlestam variables are $\hhs = (p_g + p_q)^2/Q^2$
and $\hht = (p_g + p_{\bar{q}})^2/Q^2$ with $\hhs + \hht +\hhu = 1$.

Now, for 3-parton events the thrust variable $\tau=1-T$ reduces to
\begin{equation}
  \tau =  \min (s, t, u) .
\end{equation}
So that
\begin{eqnarray}
\left[  \frac{1}{\sigma_0} \frac{\rd \sigma}{\rd \tau}\right]_{\mathrm{parton}}
&=& \delta (\tau) +
  \alp \left[ \frac{2 (3 \tau^2 - 3 \tau + 2)}{\tau (1 - \tau)} \log
  \frac{1 - 2\tau}{\tau} - \frac{3 (1 - 3 \tau) (1 + \tau)}{\tau} \right]\\
  &=& \delta (\tau) + \alp \left[ \frac{- 4 \log \tau - 3}{\tau} \right] +
  \alp d_{\mathrm{fin}} (\tau). \label{partondc}
\end{eqnarray}
The second line is written to manifest the singularity structure. The
remainder $d_{\mathrm{fin}} (\tau)$ is finite as $\tau \rightarrow 0$.

Instead of the differential thrust distribution, it is useful to work directly
with the integrated quantity
\begin{equation}
  R (\tau) = \int_0^{\tau} \frac{d \sigma}{d \tau'} d \tau' .
\end{equation}
For small $\tau$, 
\begin{equation}
  R_{\mathrm{parton}} (\tau) \sim - 2 \alp \log^2 \tau - 3 \alp \log\tau .
\end{equation}
Here we see explicitly the large logarithms $\alpha_s \log^2 \tau$ and 
$\alpha_s\log \tau$ which demand resummation.

A useful theoretical trick, used in~\cite{Catani:1992ua} and in~\cite{Fleming:2007qr}, 
is to employ a hemisphere
mass definition. This greatly simplifies the factorization formula in the
two-jet limit. The hemisphere momentum $p_L (p_R)$ is defined to be the sum of
the 4-momenta of all the particles in the hemisphere of the left (right) jet.
Then thrust reduces to $\tau = (p_L^2 + p_R^2) / Q^2$, and we can calculate it
from
\begin{equation}
  \frac{\rd \sigma}{\rd \tau} 
= \int \frac{\rd^2 \sigma}{\rd p_L^2\, \rd p_R^2} \delta (\tau
  - \frac{p_L^2 + p_R^2}{Q^2}) , 
\end{equation}
which we will see has a closed form expression.

The traditional approach to resummation which we review here is due to Catani 
et al.~\cite{Catani:1992ua}. The basic idea is that terms of the form $(\alp \log^2 \tau)^n$ come
from multiple real collinear and soft emissions. The kinematics of multiple
collinear emissions can be modeled by parton branchings, with branching
probabilities proportional to the Altarelli-Parisi splitting functions
\begin{equation}
  \frac{\rd^2 \sigma^{n + 1}}{\rd t\, \rd z} 
= \rd \sigma^n \times P_{q q} = \rd \sigma^n
  \times \alp \frac{1}{t} \frac{1 + z^2}{1 - z} .
\end{equation}
In the two-jet limit, the corrections due to soft effects can be modeled by
imposing angular ordering. All of this follows from the coherent-branching
algorithm developed in~\cite{Catani:1990rr}.
In fact, more general methods are available for handling soft emissions 
(for example, ~\cite{Catani:1996vz}). 
But, to NLO for event shapes like thrust, the methods set up in~\cite{Catani:1992ua} are
sufficient to compare to SCET.

In the two-jet limit, the factorized expression for the differential cross
section in CTTW is
\begin{equation}
\left[\frac{1}{\sigma_0}  \frac{\rd^2 \sigma_2}{\rd p_L^2\, 
\rd p_R^2}\right]_\mathrm{CTTW} = \JC (Q^2, p_L^2) \JC (Q^2, p_R^2). 
\label{factQCD} 
\end{equation}
The subscript on $\sigma_2$ refers to the 2-jet contribution.
$\JC$ has a simple physical interpretation: it is the probability
of finding a final state jet with invariant mass $p^2$. Thus, it satisfies
\begin{equation}
  \int_0^{\infty} \rd p^2 \JC(Q^2, p^2) = 1 .
\end{equation}
$Q$ is a scale associated with the hard process. We will see later that the equivalent
jet function in SCET has a different normalization.

At leading order the jet is a massless parton and so $\JC (Q^2,
p^2) = \delta (p^2)$. At next-to-leading order, the coherent branching
algorithm allows one-angular ordered emission. So
\begin{equation}
  \JC(Q^2, p^2) = \delta (p^2) +  \alp \int_0^{Q^2} \frac{\rd
  \tilde{q}^2}{\tilde{q}^2} \int_0^1\rd z 
\frac{1 + z^2}{1 - z} \delta (p^2 - z (1
  - z) \tilde{q}^2) + \cdots . \label{qcdz}
\end{equation}
Here $z$ is the energy fraction in the emission and $\tilde{q}^2 =
\frac{p^2}{z (1 - z)} \approx E^2 (1 - \cos \theta)$. The angular ordering
constraint for the coherence of soft emissions is implicit in the restriction
$\tilde{q} < Q$; for $p^2>0$ it cuts off the infrared divergences in the integral as $z\to1$.
Evaluating~\eqref{qcdz} gives
\begin{equation}
  \JC(Q^2, p^2) = \delta (p^2) + \alp \left[ \frac{- 2
  \log \frac{p^2}{Q^2} - \frac{3}{2}}{p^2} \right]_{\star}^{[p^2, Q^2]} + \cdots .
 \label{Jfirst}
\end{equation}
Thus, to first order in $\alpha_s$,
\begin{equation}
\left[\frac{1}{\sigma_0} \frac{\rd \sigma_2}{\rd \tau}\right]_\mathrm{CTTW} = \delta (\tau) + \alp \left[ \frac{- 4 \log \tau
  - 3}{\tau} \right]_{\star}^{[\tau, 1]},  \label{fixedR}
\end{equation}
which reproduces the divergent as $\tau\to0$ part of the the parton 
model result~\eqref{partondc}.

Here we have introduced the $\star$ distribution (alternatively called the R-
or $\mu$-distribution), which is a generalization of the +-distribution. It is
uniquely defined by the two conditions
\begin{equation}
  \text{$[f (x)]^{[x, a]}_{\star} = f (x)$ for $x > 0$}
\end{equation}
\begin{equation}
  \int_0^a d x [f (x)]^{[x, a]}_{\star} g (x) = \int_0^a d x f (x) \left[ g
  (x) - g (0) \right]
\end{equation}
For clarity, we have added to the notation an explicit instance of the
dependent variable $x$ in $[f]_{\star}^{[x, a]}$. A useful relation is
\begin{equation}
  [f (x)]^{[x, a]}_{\star} = [f (x)]^{[x, b]}_{\star} 
+ \delta (x) \int_a^b d  x' f (x') .
\end{equation}
This will be used extensively in the next section.

The leading order resummation is preformed by iterating the angular-ordered
emissions. This leads to an integro-differential equation for
$\JC(Q^2,p^2)$, similar to the evolution equation for
parton-distribution functions. The equation is solved in an integrated form,
and the resummed expression for the integrated thrust in the two jet limit is
\begin{equation}
\left[\frac{1}{\sigma_0}  R_2 (\tau)\right]_\mathrm{CTTW} =
\exp \left [ - 2 \alp \log^2 \tau - 3\alp \log \tau  \right ] 
\frac{e^{- 2 \gamma_E \eta}}{\Gamma [2 \eta+1]},  \label{qcdresum}
\end{equation}
where
\begin{equation}
  \eta = -2 \alp \log \tau  . \label{etaqcd}
\end{equation}
This equation is a combination of expressions in~\cite{Catani:1992ua} taken 
with $\beta = 0$. Expanding $R_2'(\tau)=\frac{\rd \sigma_2}{\rd \tau}$ 
to order $\alpha_s$ the fixed order result~\eqref{fixedR} is reproduced. 
This
expression is resummed in the sense that 
$R_2' (\tau) \rightarrow 0$ as $\tau\to0$, 
in contrast to the fixed order 
result~\eqref{fixedR} for $\rd \sigma$ or the parton
model expression~\eqref{partondc} which diverge as $\tau \rightarrow 0$.

\section{Soft-Collinear Effective Theory}

Having reviewed the way the resummed thrust distribution is traditionally
calculated, we now turn to the equivalent calculation in SCET. We will
see that there are a number of advantages of this effective field theory
treatment. Instead of a classical treatment, where multiple real emissions
are summed at the level of the cross section, the entire resummation is done
in SCET through the renormalization group. This makes explicit the various
scales in the problem, and allows greater freedom to choose those scales to
minimize the large logarithms in the distributions of interest.

The idea behind SCET is to separate out the quarks and gluons of QCD into soft
and collinear degrees of freedom. A collinear field is associated with a
light-like direction $n^{\mu}$. The component of its momentum in that direction,
$p^- = \bar{n} \cdot p$, must be much larger than any of the other momentum
components. All the QCD degrees of freedom which could change this momentum
have been integrated out, so $p^-$ appears as a fixed label. For example, a
collinear quark is written as $\xi_n (p)$ with a fixed $p^-$. In addition to
collinear quarks and gluons, there are soft quarks and gluons. These soft
fields can only interact with each other or transfer momentum to the soft
components $p^+ = n \cdot p$ of a collinear field.

The Lagrangian of SCET has a separate gauge invariance associated with soft
gluons and gluons in each collinear direction. It is useful to maintain this
gauge invariance explicitly in the operators of the theory with the use of
Wilson lines. For example, a two-jet operator is
\begin{equation}
  \mathcal{O}_2 = \bar{\xi}_{\bar{n}}W_{\bar{n}} Y_{\bar{n}}\gamma^{\mu}
  Y_n^{\dag}   W_n^{\dag} \xi_n ,
\end{equation}
where
\begin{equation}
  W_n (x) = P \exp \left\{ i g\int \rd s \bar{n} \cdot A_n ( \bar{n}
  s + x) \right\}
\end{equation}
\begin{equation}
  Y_n (x) = P \exp \left\{ i g \int \rd s n \cdot A_s (n s + x)
  \right\}.
\end{equation}
The collinear Wilson lines $W_n$ maintain collinear gauge invariance and
the soft Wilson lines $Y_n$ maintain soft gauge invariance.

The starting point in the effective theory approach to event shapes is again
factorization in the two-jet limit. In SCET, the event-shape distributions
near the endpoint come from matrix elements of
$\mathcal{O}_2$~\cite{Fleming:2007qr,Bauer:2006qp,Bauer:2002ie,Lee:2006nr}.
In terms of the hemisphere masses $p_L$ and $p_R$ defined above,
factorization implies~\cite{Fleming:2007qr}
\begin{equation}
\left[  \frac{\rd^2 \sigma_2}{\rd p_L^2\, \rd p_R^2} 
\right]_\mathrm{SCET}= |C_H (\mu)|^2 \int \rd k_L \rd k_R J
  (p_L^2 - Q k_L, \mu) J (p_R^2 - Q k_R, \mu) S (k_L, k_R, \mu) .
  \label{factorform}
\end{equation}
The hard function $C_H(\mu)$, the jet functions $J(p^2,\mu)$ and the soft function 
$S(k,\mu)$ will be defined below.

The form of this factorized expression
\eqref{factorform} has a physical explanation. Each jet function $J (p^2, \mu)$
comes from one of the collinear quarks. It represents, like $\JC$,
the probability for producing a jet of invariant mass $p^2$ from that
collinear field (the precise relation to $\JC$ is explored below).
Recall that the large component of the momentum of a collinear field 
$p^- = \bar{n}\cdot p$ is fixed. Since, in the two-jet limit, the jets are back to back with
center of mass energy $Q$, we must have $p^- \sim Q$. The small component $p^+
= n \cdot p$ can vary. If there are no soft effects, then the hemisphere mass
is simply $p_H^2 = p^2 = p^- p^+$. However, as the factorization formula
implies, the collinear jet can give up some soft momentum to the soft QCD
background, leading to $p^+ \rightarrow p^+ - k$. The hemisphere mass is
unchanged by this emission, but now $p^2 = p^- (p^+ - k) = p^2_H - Q k$, which
explains the form of \eqref{factorform}.

The function $C_H$ is a hard function, which comes from integrating out hard
modes of QCD in matching to SCET. Demanding
\begin{equation}
  \left\langle \bar{q} \gamma^{\mu} q \right\rangle = C_H \left\langle \xi_n
  \gamma^{\mu} \xi_{\bar{n}} \right\rangle
\end{equation}
for states with two external quarks lets us calculate $C_H$ order by order in
perturbation theory. The computation entails taking the difference between QCD
and SCET graphs, such as
\begin{equation}
C_H \sim
\SetScale{0.5}
\fcolorbox{white}{white}{
  \begin{picture}(60,20) (0,0)
    \SetWidth{0.5}
    \SetColor{Black}
    \LongArrow(45,24)(0,54)
    \Photon(45,24)(45,-21){4}{3}
    \LongArrow(45,24)(90,54)
    \GlueArc(44.69,30.91)(24.09,160.38,22.16){-4.5}{4.29}
  \end{picture}
}
-
\SetScale{0.3}
\fcolorbox{white}{white}{
  \begin{picture}(60,20) (0,5)
    \SetWidth{0.5}
    \SetColor{Black}
    \CArc(117,18)(20.62,129,489)
    \LongArrow(134,32)(230,111)
    \LongArrow(98,29)(4,111)
    \Line(103,4)(130,31)
    \CArc(169.48,38.68)(57.48,59.09,180.68)
    \GlueArc(167.18,42.56)(50.25,182.92,53.6){-7.5}{6.67}
    \Line(132,3)(103,31)
  \end{picture}
},
\end{equation}
where the $\otimes$ refers to an insertion of ${\mathcal O}_2$,
and the right diagram is only a representative contribution (for
details, see~\cite{Bauer:2006mk}).
The difference is finite because the infrared divergences in
QCD and SCET are the same and the UV divergences are removed
with counterterms. At one-loop the matching
gives~\cite{Manohar:2003vb,Bauer:2006mk}
\begin{equation}
  C_H (\mu) = 1 + c_H + 
\frac{\Gamma_H}{2} \log^2 \frac{\mu^2}{Q^2} 
- \gamma_H \log \frac{\mu^2}{Q^2}, \label{hardfixed}
\end{equation}
with 
$c_H = \alp \left( - 4 + \frac{7 \pi^2}{12} - \frac{3\pi}{2} i \right)$,
$\Gamma_H = - \alp$ and 
$\gamma_H = \frac{3}{2} \alp$. 

The anomalous dimension
and renormalization group evolution of $C_H(\mu)$ can be found from the UV
divergences of the same 1-loop graphs. The RG equation is
\begin{equation}
  \frac{\rd C_H (\mu)}{\rd \log \mu} 
= (-2 \Gamma_H \log  \frac{Q^2}{\mu^2} -2\gamma_H) C_H(\mu) ,
\end{equation}
with solution
\begin{equation}
  C_H (\mu) 
= C_H (\mu_h) \exp \left[ \frac{\Gamma_H}{2} \log^2  \frac{\mu^2}{\mu_h^2} 
-\gamma_H \log \frac{\mu^2}{\mu_h^2} \right]
\left( \frac{Q^2}{\mu_h^2}\right)^{-\Gamma_H 
\log\frac{\mu^2}{\mu_h^2}} .
\end{equation}
To first order in $\alpha_s$, the $\mu_h$ dependence drops out and
\eqref{hardfixed} is reproduced. The natural matching scale $\mu_h=Q$ is manifest
in this expression.

The jet function $J (p^2, \mu)$ is the imaginary part of the propagator of a
collinear quark. It is defined by~\cite{Becher:2006qw,Fleming:2007qr}
\begin{equation}
  J (p^2, \mu) = - \frac{1}{\pi ( \bar{n} \cdot p)} \mathrm{Im} \left[ \int d^4
  x e^{- i p x} \left\langle T 
\left\{ (\bar{\xi}_{\bar{n}} W_{\bar{n}}^\dagger) (0)
  \not\!{\bar{n}} (W_n \xi_n) (x) \right\} \right\rangle \right] .
\end{equation}
It can be thought of as the spectral density for a jet of collimated
particles interacting with a soft QCD background. The jet function can be
calculated order-by-order in perturbation theory through the discontinuity of
conventional Feynman diagrams, such as
\begin{equation}
J(p^2,\mu) \sim \mathrm{Disc} \left\{
\SetScale{0.3}
\fcolorbox{white}{white}{
  \begin{picture}(80,20) (0,3)
    \SetWidth{0.5}
    \SetColor{Black}
    \CArc(91.53,16.08)(50.48,-1.23,164)
    \GlueArc(90.8,13.32)(47.23,158.02,2.04){-7.5}{7.76}
   \CArc(27,16)(20.62,129,489)
    \Line(13,2)(40,29)
    \Line(42,1)(13,29)
    \CArc(258,16)(20.62,129,489)
    \Line(244,2)(271,29)
    \Line(273,1)(244,29)
    \Line(49,15)(237,15)
  \end{picture}
}
+
\fcolorbox{white}{white}{
  \begin{picture}(80,20)(0,6)
    \SetWidth{0.5}
    \SetColor{Black}
    \CArc(27,27)(20.62,129,489)
    \Line(13,13)(40,40)
    \Line(42,12)(13,40)
    \CArc(258,27)(20.62,129,489)
    \Line(244,13)(271,40)
    \Line(273,12)(244,40)
    \Line(49,26)(237,26)
    \CArc(143.11,-67.94)(147.96,48.58,132.58)
    \GlueArc(143.13,-85.74)(160.74,128.53,52.04){-7.5}{13.9}
  \end{picture}
}
+
\fcolorbox{white}{white}{
  \begin{picture}(80,20) (0,3)
    \SetWidth{0.5}
    \SetColor{Black}
    \CArc(27,20)(20.62,129,489)
    \Line(13,6)(40,33)
    \Line(42,5)(13,33)
    \CArc(258,20)(20.62,129,489)
    \Line(244,6)(271,33)
    \Line(273,5)(244,33)
    \Line(49,19)(237,19)
    \GlueArc(138.5,16.64)(51.55,177.38,2.62){-7.5}{9.8}
    \CArc(139.5,17.76)(55.51,1.28,178.72)
  \end{picture}
}
+\cdots
\right\} .
\end{equation}
Some of the infrared divergences in these graphs are cut off
because the collinear
quark is taken to have invariant mass $p^2$.
These calculations have been done in the context of 
$b \rightarrow s \gamma$~\cite{Bauer:2003pi,Bosch:2004th}
and for deep-inelastic scattering~\cite{Manohar:2003vb,Becher:2006mr}
and the jet function is known to two loops~\cite{Becher:2006qw}. 
Due to factorization, the same jet function applies in these processes and
in $e^+ e^-$ annihilation. To first order in $\alpha_s$, the result
is
\begin{equation}
  J (p^2, \mu) = \delta (p^2) \left[ 1 + c_J \right] + \left[ \frac{\Gamma_J
  \log \frac{p^2}{\mu^2} + \gamma_J}{p^2} \right]^{[p^2, \mu^2]}_{\star},
 \label{jetfixed}
\end{equation}
with $c_J = \alp \left( \frac{7}{2} - \frac{\pi^2}{2} \right)$, $\Gamma_J =
2 \alp$ and $\gamma_J = - \frac{3}{2} \alp$.

The renormalization group evolution
of the jet function, in contrast to that of the
hard function, is non-local in $p^2$
\begin{equation}
  \frac{\rd J (p^2, \mu)}{\rd \log \mu} = \left[-2 \Gamma_J \log
  \frac{p^2}{\mu^2} -2 \gamma_J \right] J (p^2, \mu) + 2\Gamma_J \int_0^{p^2} d
  q^2 \frac{J (p^2, \mu) - J (q^2, \mu)}{p^2 - q^2} .
\end{equation}
This is similar to the Altarelli-Parisi equation for the evolution of the
parton distribution functions (pdfs). In contrast to pdfs, however, the jet
function $J (p^2, \mu)$ is perturbative as long as 
$p^2 >\lqcd$. Simplification is achieved through use of the Laplace
transform~\cite{Becher:2006mr},
\begin{equation}
  \tilde{j} (\nu) = \int d p^2 e^{- \nu p^2} J (p^2, \mu) ,
\end{equation}
whereby the evolution becomes local in $\nu$
\begin{equation}
  \frac{\rd \tilde{j} (\nu, \mu)}{\rd \log \mu} = \left(-2 \Gamma_J \log
  \frac{1}{\mu^2 \nu e^{\gamma_E}} - 2\gamma_J \right) \tilde{j} (\nu, \mu) .
\end{equation}
Now the RGE can be solved like that of $C_H$:
\begin{equation}
  \tilde{j}(\nu,\mu) = \exp \left[ \frac{\Gamma_J}{2} \log^2 \frac{\mu^2}{\mu_j^2} 
- \gamma_J  \log \frac{\mu^2}{\mu_j^2} \right]  \left( \nu e^{\gamma_E} \mu_j^2
  \right)^{-\eta_j} \tilde{j}(\nu,\mu_j)
\end{equation}
with
\begin{equation}
  \eta_j = -\Gamma_J \log \frac{\mu^2}{\mu_j^2} .
\end{equation}
 Finally, the inverse Laplace transform produces~\cite{Becher:2006mr}
\begin{equation}
  J (p^2, \mu) = \exp \left[ \frac{\Gamma_J}{2} \log^2 \frac{\mu^2}{\mu_j^2} -
  \gamma_J \log \frac{\mu^2}{\mu_j^2} \right] \tilde{j} \left(
  \partial_{\eta_j} \right) \left[ \frac{1}{p^2} \left( \frac{p^2}{\mu_j^2}
  \right)^{\eta_j} \right]_{\star}^{[p^2, \mu_j^2]} \frac{e^{- \gamma_E
  \eta_j}}{\Gamma [\eta_j]},
\end{equation}
where
\begin{equation}
  \tilde{j} \left( \partial_{\eta_j} \right) = 1 + c_J + \Gamma_J
  \frac{\pi^2}{12} + \frac{\Gamma_J}{2} \partial_{\eta_j}^2 + \gamma_J
  \partial_{\eta_j}.
\end{equation}
The functional dependence on $\partial_{\eta_j}$ in $\tilde{j}(\partial_{\eta_j})$ comes
from a functional dependence on $\log\frac{p^2}{\mu^2}$ in the
fixed order expression for the jet function~\eqref{jetfixed} 
at the matching scale $\mu=\mu_j$.

Finally, we have to calculate the soft function $S (k_L, k_R, \mu)$. It is
defined though matrix elements of soft Wilson 
lines~\cite{Korchemsky:1993uz,Fleming:2007qr}
\begin{equation}
  S (k_L, k_R,\mu) =\sum_X
\langle 0|Y^\star_{\bar{n}} Y_n| X\rangle
\langle X|Y^\dagger_n Y_{\bar{n}}^{\dagger\star} |0 \rangle
\label{softdef} .
\end{equation}
Unlike the jet function, the soft function is often evaluated 
at scales $k \sim\lqcd$, 
where it is non-perturbative. It can, in
principle, be measured experimentally and then be evolved perturbatively,
like the pdfs. 
Or it can be modeled~\cite{Korchemsky:2000kp,Korchemsky:1999kt,Fleming:2007qr}.
However, in our
case, since we are taking $\alpha_s$ to be small and fixed, it can be
calculated in perturbation theory. Even if $S$ is non-perturbative, the
perturbative calculation is useful because the UV divergences dictate the
anomalous dimensions and hence the evolution equation.

The soft function, 
being completely defined in terms of Wilson lines, can be calculated in QCD
through diagrams such as
\vspace{-0.5cm}
\begin{equation}
S(k_L,k_R,\mu) \sim 
\SetScale{0.3}
\fcolorbox{white}{white}{
  \begin{picture}(30,50)(0,20) 
   \SetWidth{0.5}
    \SetColor{Black}
    \ArrowLine(0,180)(45,90)
    \ArrowLine(45,90)(0,0)
    \ArrowLine(45,180)(90,90)
    \ArrowLine(90,90)(45,0)
    \Gluon(30,60)(60,30){6.5}{2.57}
  \end{picture}
}
+
\fcolorbox{white}{white}{
  \begin{picture}(30,50)(0,20) 
   \SetWidth{0.5}
    \SetColor{Black}
    \ArrowLine(0,180)(45,90)
    \ArrowLine(45,90)(0,0)
    \ArrowLine(45,180)(90,90)
    \ArrowLine(90,90)(45,0)
    \GlueArc(-25.3,123.3)(91.48,3.57,-58.86){-5}{6.6} 
  \end{picture}
}
+
\fcolorbox{white}{white}{
  \begin{picture}(30,50)(0,20) 
   \SetWidth{0.5}
    \SetColor{Black}
     \SetColor{Black}
    \ArrowLine(0,180)(45,90)
    \ArrowLine(45,90)(0,0)
    \ArrowLine(45,180)(90,90)
    \ArrowLine(90,90)(45,0)
    \GlueArc(7.84,90.16)(47.33,71.32,-72.59){-5}{8.14}
  \end{picture}
}
+\cdots ,
\end{equation}
\vspace{0.3cm}

\noindent
where the kinked lines are Wilson lines in the $n$ and $\bar{n}$ directions.
To order $\alpha_s$,  $S (k_L, k_R)$ can be written as a product
\begin{equation}
  S (k_L, k_R, \mu) = S (k_L, \mu) S (k_R, \mu)
\end{equation}
where~\cite{Korchemsky:1993uz}
\begin{equation}
  S (k, \mu) = \delta (k) \left[ 1 + c_S \right] 
+ \left[  \frac{\Gamma_S \log \frac{k}{\mu} + \gamma_S}{k} \right]^{[k, \mu]}_\star
\label{Sfirst} ,
\end{equation}
with $c_S = \alp \frac{\pi^2}{12}$, $\Gamma_S = - 4 \alp$ and $\gamma_S =
0$.

The evolution of the soft function involves the same kind of non-local
equation as for the jet function. This can be seen from examining
its anomalous dimension directly, or from RG invariance of the convolution
appearing in the factorization formula~\eqref{factorform}.
The solution is therefore similar. Explicitly,
\begin{equation}
  S (k, \mu) = \exp \left[ \frac{\Gamma_S}{2} \log^2 \frac{\mu}{\mu_s} -
  \gamma_S \log \frac{\mu}{\mu_s} \right] \tilde{s} (\partial_{\eta_s}) \left[
  \frac{1}{k} \left( \frac{k}{\mu_s} \right)^{\eta_s} \right]_{\star}^{[k,
  \mu_s]} \frac{e^{- \gamma_E \eta_s}}{\Gamma [\eta_s]} ,
\end{equation}
where
\begin{equation}
  \tilde{s} (\partial_{\eta_s}) = 1 + c_S + \Gamma_S \frac{\pi^2}{12} +
  \frac{\Gamma_S}{2} \partial_{\eta_s}^2 + \gamma_S \partial_{\eta_s}
\end{equation}
and
\begin{equation}
  \eta_s = - \Gamma_S \log \frac{\mu}{\mu_s}.
\end{equation}
Expanding to first order in $\alpha_s$ reproduces~\eqref{Sfirst}.

Now we have all the ingredients appearing in the
 the factorization formula~\eqref{factorform}.
At order $\alpha_s$, the result is
\begin{equation}
\left[\frac{1}{\sigma_0}  \frac{\rd^2 \sigma_2}{\rd p_L^2\, \rd p_R^2} \right]_\mathrm{SCET}
= \delta (p^2_L) \delta (p_R^2) \left[  
1 + \alp \left( - 1 + \frac{\pi^2}{3} \right) 
\right] \nonumber
\end{equation}
\begin{equation}
 +\alp \delta (p_R^2) 
\left[ \frac{2 \log \frac{Q^2}{p^2_L} 
-  \frac{3}{2}}{p^2_L} \right]^{[p^2_L, Q^2]}_\star 
+ \alp \delta (p_L^2) 
\left[ \frac{2 \log \frac{Q^2}{p^2_R} 
-  \frac{3}{2}}{p^2_R} \right]^{[p^2_R, Q^2]}_\star .
\label{dplpr}
\end{equation}
Note that the dependence on all the scales $\mu, \mu_h, \mu_j$ and $\mu_s$ has
completely canceled. This is a non-trivial result which requires three relations
among the six anomalous dimensions:
\begin{equation}
\Gamma_J+\frac{\Gamma_S}{2}=\Gamma_H+\Gamma_J+\frac{\Gamma_S}{4}
=\gamma_H + \gamma_J+\frac{\gamma_S}{2} = 0 
\end{equation}
and is a strong consistency check on the entire formalism.~\footnote{
The relations among the $\Gamma$s can be understood on more general
grounds by relating the $\Gamma$s to a universal cusp 
anomalous dimension~\cite{Korchemsky:1987wg}.}

Already SCET can be compared to CTTW. The differential
distribution in SCET, Eq.~\eqref{dplpr}, matches 
the CTTW distribution, Eq.~\eqref{factQCD},
when the jet functions are expanded to 
first order using Eq.~\eqref{Jfirst}.
The SCET expression has an
additional finite piece (the $- 1 + \frac{\pi^2}{3}$ term) which comes from
loop graphs which do not enter into the traditional formulation. 

Still working at leading order in $\alpha_s$, 
SCET produces a simple form for the thrust distribution near $\tau=0$:
\begin{equation}
\left[ \frac{1}{\sigma_0} \frac{\rd \sigma_2}{\rd \tau} \right]_\mathrm{SCET}
= \delta (\tau) \left[ 1 + \alp \left( - 1 + \frac{\pi^2}{3} \right) \right] 
+ \alp \left[ \frac{- 4 \log \tau - 3}{\tau} \right]^{[\tau, 1]}_\star .
\label{scetsig2}
\end{equation}
This reproduces the leading behavior for small $\tau$ of the 
both the parton model expression, Eq.~\eqref{partondc},
and the resummed expression 
with the functions $\JC$ expanded to first order, 
Eq.~\eqref{fixedR}. 

To get the resummed thrust distribution from SCET, 
we need to calculate a couple of
convolution integrals. First, the soft and jet functions must be combined into
\begin{equation}
  K (p^2, \mu) = \int d k J (p^2 - k Q, \mu) S (k, \mu) .
\end{equation}
Second, the $K$ functions must be integrated against the event shape. For
thrust, we need
\begin{equation}
  R_2(\tau) = | C_H(\mu)|^2 \int K (p^2_L, \mu) K (p_R^2, \mu) \theta (\tau -
  \frac{p^2_L + p_R^2}{Q^2}).
\end{equation}
Both of these convolutions can be evaluated by performing the Laplace
transform. 

Note that the function names can be misleading. 
Here, $C_H (\mu) K (p^2, \mu)$ 
(and not just the SCET jet function $J (p^2, \mu)$) 
is playing the
role of $\JC(p^2)$ from the CTTW formulation. Thus, the
CTTW jet function is a combination of the soft and jet functions in
SCET, as expected because the coherent branching algorithm used to derive it
incorporates both soft and collinear effects.

Solving the $K$ function with the Laplace transform techniques gives
\begin{equation}
  K (p^2, \mu) = \exp \left[ \frac{\Gamma_J}{2} \log^2 \frac{\mu^2}{\mu_j^2} -
  \gamma_J \log \frac{\mu^2}{\mu_j^2} \right] \exp \left[ \frac{\Gamma_S}{2}
  \log^2 \frac{\mu}{\mu_s} - \gamma_S \log \frac{\mu}{\mu_s} \right] \left(
  \frac{Q \mu_s}{\mu_j^2} \right)^{-\eta_s}
\end{equation}
\begin{equation}
  \times \left[ \frac{\Gamma_S}{2} \log^2 \frac{Q \mu_s}{\mu_j^2} - \Gamma_S
  \log \frac{Q \mu_s}{\mu_j^2} \partial_{\eta_k} + \tilde{k}
  (\partial_{\eta_k}) \right] \left[ \frac{1}{p^2} \left( \frac{p^2}{\mu_j^2}
  \right)^{\eta_k} \right]_{\star}^{[p^2, \mu_j^2]} \frac{e^{- \gamma_E
  \eta_k}}{\Gamma [\eta_k]},
 \label{resid}
\end{equation}
where
\begin{equation}
  \tilde{k} (\partial_{\eta_k}) = 1 + c_J + c_S + \frac{\pi^2}{12} \left(
  \Gamma_J + \Gamma_S \right) + \left( \frac{\Gamma_J + \Gamma_S}{2} \right)
  \partial_{\eta_k}^2 + \left( \gamma_J + \gamma_S \right)
  \partial_{\eta_k}
\end{equation}
and
\begin{equation}
  \eta_k = \eta_s + \eta_j = 2 \alp \log \frac{\mu_j^2}{\mu_s^2} .
\end{equation}
Note the residual logarithms of $Q \mu_s/\mu_j^2$ in line 
\eqref{resid} which are not resummed.
These imply that we cannot choose $\mu_j = \mu_s$. Instead, 
the natural scales (those which remove the large logs) 
should satisfy
 $\mu_s = \mu_j^2 / Q$. Then, we can evolve either the jet
function from $\mu_j$ down to $\mu_s$ or the soft function from $\mu_s$ up to
$\mu_j$ but there is not a single scale which can minimize all the large logs.

Finally, to evaluate the convolution for the thrust integral, we use the
Laplace transform of the $\theta$-function
\begin{equation}
\int_0^\infty \rd p^2 \theta(p^2) e^{-\nu p^2}=\frac{1}{\nu} .
\end{equation}
Combining this with the expression for $K (p^2, \mu)$ above,
the SCET prediction for thrust in the 2-jet limit is
\begin{equation}
\left[\frac{1}{\sigma_0}  R_2 (\tau)\right]_\mathrm{SCET} 
= |C_H (\mu_h) |^2 \exp \left[ 2 \alp \log^2
  \frac{\mu_h^2}{\mu_j^2} + 3 \alp \log \frac{\mu_h^2}{\mu_j^2} - \alp
  \log^2 \frac{\mu_h^2}{\mu_s^2} \right] \left( \frac{Q \mu_s}{\mu_j^2}
  \right)^{- 4 \alp \log \frac{\mu_h^2}{\mu_s^2}} \nonumber
\end{equation}
\begin{equation}
  \times \left[ - 2 \alp \log^2 \frac{Q \mu_s}{\mu_j^2} + 4 \alp \log
  \frac{Q \mu_s}{\mu_j^2}\partial_{2\eta_k} + \tilde{k} (\partial_{2\eta_k}) \right]^2 \left(
  \frac{Q^2 \tau}{\mu_j^2} \right)^{2 \eta_k} \frac{e^{- 2 \gamma_E
  \eta_k}}{\Gamma [2 \eta_k + 1]} .
\end{equation}
Note that this expression is explicitly independent of $\mu$.

Now, choosing the natural
scales $\mu_h = Q$, 
$\mu_j = Q \sqrt{ \tau}$, and
$\mu_s = \mu_j^2/Q=Q\tau$ 
to remove the large logs, the thrust distribution becomes
\begin{equation}
\left[\frac{1}{\sigma_0}  R_2 (\tau)\right]_\mathrm{SCET} 
 = \exp \left [ - 2 \alp \log^2 \tau - 3 \alp \log \tau
  \right ] \tilde{r} (\partial_{\eta}) \frac{e^{- 2 \gamma_E \eta}}{\Gamma
  [2 \eta + 1]}
\end{equation}
with
\begin{equation}
  \tilde{r} (\partial_{\eta_{}}) = 1 + (- 1 + \frac{\pi^2}{3}) \alp -
  \frac{\pi^2}{3} \alp - \frac{1}{2} \alp \partial_{\eta}^2 
- \frac{3}{2} \alp
  \partial_{\eta}  \label{rtwiddef}
\end{equation}
and $\eta = - 2 \alp \log \tau$ as in Eq.~\eqref{etaqcd}.

This expression would be identical to the CTTW expression, 
Eq.~\eqref{qcdresum}, if $\tilde{r} = 1$. 
Recall that the function $\tilde{r}$ comes from
1-loop matching in SCET, 
which turns into boundary conditions for the renormalization group evolution. 
In the approach of CTTW the boundary condition is simply that 
$\JC(p^2) = \delta(p^2)$. 
Nevertheless, the effect of $\tilde{r}\ne 1$ 
to order
$\alpha_s$ is only to provide a finite constant,
as can been seen from comparing equations~\eqref{scetsig2} and~\eqref{fixedR}.
That is, there is no contribution of order $\eta$ in the difference,
only of order $\eta^2$ and higher. Beyond order $\alpha_s$,
the $\partial_{\eta}$ does change the $\eta$ dependence. But since
$\eta \sim \alpha_s \log \tau$ 
these terms are subleading to the dominant $\alpha_s
\log^2 \tau$ and can get corrections from higher-loop effects. For example,
at two-loops, a term of the form $\alpha_s^2 \log^2 \tau$ is generated. Thus,
SCET and CTTW agree to first order in $\alpha_s$ and for
the resummation of the leading large logarithms, which is
the order to we have been working.

\section{Matching to hard emissions}
Now let us look at how fixed order results and resummation are combined. There
are two elements to this: (1) matching to hard emissions to get the
differential distribution correct away from the two-jet region (i.e. away from
$\tau = 0$); and (2) including finite parts of loops to reproduce fixed order
inclusive results.

Let us begin with the matching procedure
described in~\cite{Catani:1992ua}. To match to
the hard emissions at order $\alpha_s$ we need the parton model differential
cross section from equation (\ref{partondc}). The divergent part is already
contained in the two jet contribution, so the remainder is
\begin{equation}
  \Df (\tau) = \int_0^{\tau} d \tau' d_{\mathrm{fin}} (\tau') 
\label{Deq}
\end{equation}
\begin{equation} 
\nonumber
= \log \tau[6\tau + 4  \log (1 - \tau)]
- 2 \log^2 (1 - \tau) 
  + 3 (1 - 2 \tau) \log (1 - 2 \tau) 
- 4 \mathrm{Li}_2 (  \frac{\tau}{1 - \tau}) 
+6\tau +  \frac{9}{2} \tau^2 .
\end{equation}
The final result for the matched integrated thrust 
distribution from~\cite{Catani:1992ua} 
is
\begin{equation}
  \left[ \frac{1}{\sigma_0} R (\tau) \right]_{\mathrm{CTTW}} 
= (1 + \sigma_1)  \left\{ 
\left( 1 + C_1 \right) \exp \left[ - 2 \alp \log^2 \tau -
  3\alp \log \tau \right ] \frac{e^{- 2\gamma_E \eta}}{\Gamma [ 2\eta+1]}
  + \alp \Df (\tau) \right\},
 \label{matchedQCD}
\end{equation}
where $\sigma_1 = \frac{3}{2} \alp = \frac{\alpha_s}{\pi}$ is
the NLO contribution to the total cross section and $C_1 = \alp
\left( \frac{\pi^2}{3} - \frac{5}{2} \right)$
is chosen so that
$R (\frac{1}{3}) = 1 + \sigma_1$ 
(recall that $\tau < \frac{1}{3}$ at order $\alpha_s$).
In the QCD approach the factor $\sigma_1=\frac{\alpha_s}{\pi}$ 
must be be determined by a
independent calculation of the total rate. This involves combining the real and
virtual contributions at order $\alpha_s$ using an appropriate infrared
regulator.
As emphasized in~\cite{Catani:1992ua} there is arbitrariness in
the matching because the hard emissions (in $\Df(\tau)$) are fixed order but
the two-jet contribution is resummed. For example, to the same accuracy $D
(\tau)$ could be multiplied by the exponential.

Now, let us turn to the matching in SCET. To include thrust distributions
away from the endpoint, we can either attempt to add power corrections to
SCET, or we can match to higher order operators as described 
in~\cite{Bauer:2006mk,Bauer:2006qp}. 
Matching
is much simpler. To perform the matching, we add new operators
\begin{equation}
  \mathcal{O}_3 = {\bar{\xi}}_{n_1} A_{n_2}^{\nu} \gamma^\mu \xi_{n_3} + \cdots,
\end{equation}
where the $\cdots$ are the additional terms coming from Wilson lines necessary
for gauge invariance. $A^{\mu}_{n_2}$ is a collinear gluon in direction $n_2$.
The matching demands that
\begin{equation}
 \langle \mathcal{O}_2\rangle_{\mu_h}
  + \langle \mathcal{O}_3 \rangle_{\mu_h} 
=  \left\langle \mathrm{QCD} \right\rangle_{\mu_h},
\label{o23}
\end{equation}
where the subscript means the matching is done at the hard scale $\mu_h$.

There is some arbitrariness in the definition of the matrix elements in SCET
due to reparameterization invariance. The matrix element of a collinear quark
$\xi_n (p)$ on a QCD quark state $|q\rangle$ is only defined up to its
soft momentum component $n \cdot p$. Moreover, a basis for summing
over directions $n$ for ${\mathcal O}_2$ and
$n_1$, $n_2$, and $n_3$ for ${\mathcal O}_3$ must be chosen as well.
A certain convention was described in~\cite{Bauer:2006mk} for
resolving these ambiguities and others are
possible. In any case, while different
conventions may shift 
the contributions from $\mathcal{O}_2$ and $\mathcal{O}_3$
in~\eqref{o23}, 
the sum is parameterization invariant.
Thus, independently of the convention we have
\begin{equation}
[\rd \sigma]_{\mathrm{SCET}}^{\mu_h} = [\rd \sigma]_{\mathrm{parton}},
\end{equation}
where 
$[\rd \sigma]_{\mathrm{parton}}\sim \langle \mathrm{QCD} \rangle^2$ 
is the tree-level parton model cross section, as shown in
 Eq.~\eqref{sigparton}.

Now, we already know that $\left\langle \mathcal{O}_2 \right\rangle^2$ 
gives~\eqref{scetsig2} to first order in $\alpha_s$ and that
the parton model distribution at order $\alpha_s$ is~\eqref{partondc}.
Thus, with obvious implicit phase space factors,
\begin{equation}
  \left\langle \mathcal{O}_2 + \mathcal{O}_3 \right\rangle_{\mu_h}^2 -
  \left\langle \mathcal{O}_2 \right\rangle^2_{\mu_h} = d_{\mathrm{fin}}.
\end{equation}
So, at leading order, the contribution from the sum of 
$\langle \mathcal{O}_3\rangle^2$ and
the interference between $\mathcal{O}_2$ and $\mathcal{O}_3$ to the
differential cross section is unambiguous. The running of $\mathcal{O}_3$
could also be included even though it does not resum any large logs for the
event shapes under consideration. With running, the matrix elements of
$\mathcal{O}_2$ and $\mathcal{O}_3$ would appear with separate renormalization
kernels, and so the final differential cross section would end up depending on
the conventions chosen. 
The ambiguity could be resolved by a careful higher order
treatment, but for the purposes of comparing to the CTTW prediction for thrust,
we will simply not renormalize the finite terms.

Then,
\begin{equation}
  \left[ \frac{1}{\sigma_0} R (\tau) \right]_{\mathrm{SCET}} 
= \exp \left [ - 2
  \alp \log^2 \tau - 3 \alp \log \tau \right ] \tilde{r} (\partial_{\eta})
  \frac{e^{-2 \gamma_E \eta}}{\Gamma [1 + 2\eta]} + \alp \Df (\tau).
\label{matchedscet}
\end{equation}
With $\tilde{r} (\partial_{\eta})$ and $\eta$ as
in Eq.~\eqref{rtwiddef} and $\Df(\tau)$ in~\eqref{Deq}.
The total cross section in SCET is given
\begin{equation}
\left [  R ( \frac{1}{3})\right]_\mathrm{SCET} = 
\sigma_0\left[1 + \alp \left( - 1 + \frac{\pi^2}{3} \right) + \alp
  \Df ( \frac{1}{3}) \right]
=\sigma_0\left[ 1 + \frac{\alpha_s}{\pi}\right].
\end{equation}
This is the correct total $e^+ e^-$ total cross section to first order in
$\alpha_s$!

Let us review the contributions that go into the cross section. First, at
the hard scale $\mu_h=Q$, there is the finite part of the loop matching to QCD, 
$|c_H|^2
= 1+\alp (-8 + \frac{7 \pi^2}{6}) = 1+2.3 \sigma_1$. 
Next, at the
scale $\mu_j^2=p^2=Q^2\tau$ 
where we integrate out the collinear fields, the jet
functions give 
$2 c_J = \alp (7 - \pi^2) = - 1.9 \sigma_1$. At the seesaw
scale 
$\mu_s = p^2 / Q$, the soft function gives 
$2 c_S = \frac{\pi^2}{6} =
1.1 \sigma_1$, and finally the finite part of the real emission integral, away
from $\tau = 0$ gives $\sigma_3 = \alp \left( \frac{5}{2} - \frac{\pi^2}{3}\right)
= - 0.5 \sigma_1$. 
In producing the total cross section, only the soft
and jet contributions are infrared divergent. However, their convolution, which appears
in the function $K(p^2,\mu)$ is infrared finite. The hard matching and
the hard emissions are IR finite by themselves. Thus, the total cross section can
be understood as a combination of a process dependent IR finite hard part and
universal but IR regulator dependent 
contributions from soft and collinear emissions.

\begin{figure}[t]
\begin{center}
\includegraphics[width=0.95\textwidth]{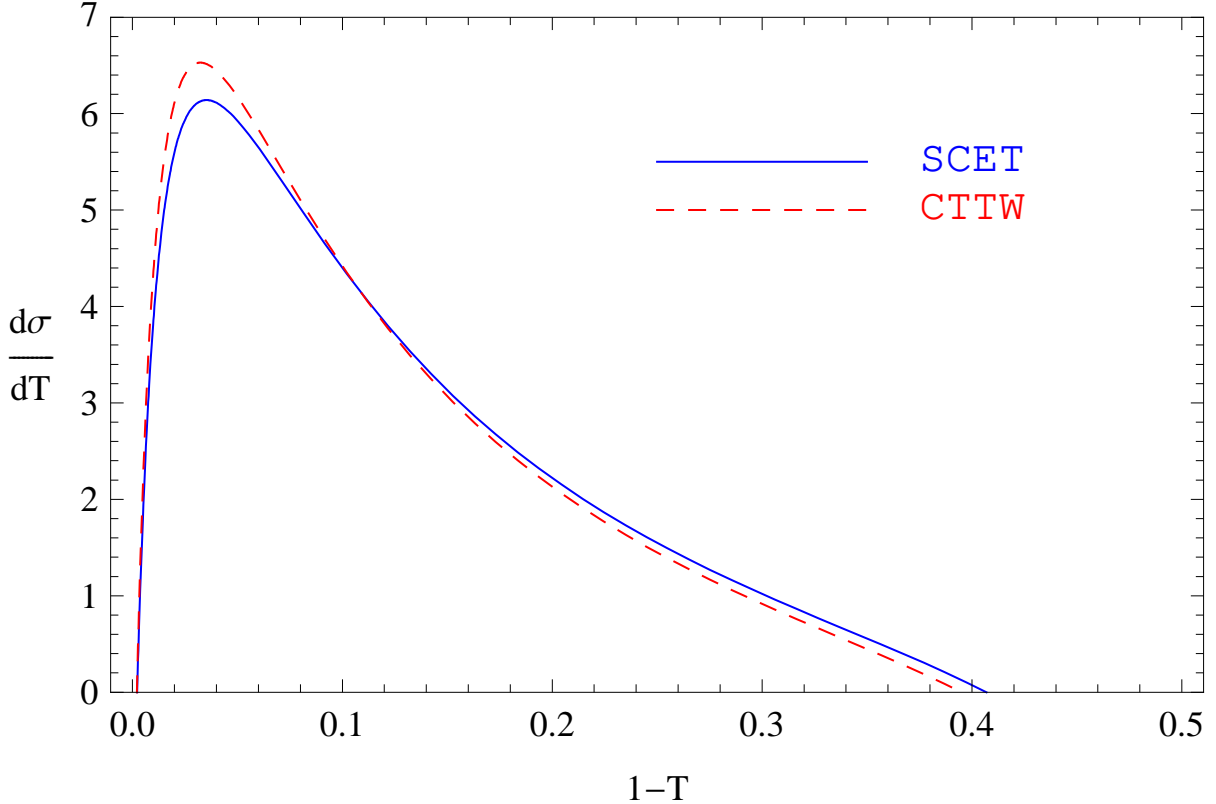} 
\vspace{0.0cm}
\caption{\label{fig:dt}
Matched resummed differential thrust distribution in SCET and in the standard
approach of CTTW with fixed coupling $\alpha_s=0.4$}
\end{center}
\end{figure}

The SCET thrust distribution~\eqref{matchedscet}
is compared to the CTTW thrust distribution~\eqref{matchedQCD}
in Figure~\ref{fig:dt}.
The plot is of $\frac{\rd \sigma}{\rd \tau} = R'(\tau)$ 
with $\alpha_s$ fixed at $0.4$.
A more careful rendition of the differential
thrust distribution would take the derivative of $R(\tau)$ before 
assigning the matching scale $\mu_j=Q\sqrt{\tau}$, as is done below. 
However, the effect is higher
order, and so we just plot $R'(\tau)$ directly.

\section{Generalizations \label{sec:gen}}
In this section, some simple generalizations are described. The above results
were derived assuming $\alpha_s$ to be constant in order to emphasize the 
resummation of Sudakov logarithms in contrast to large logarithms associated with the
scale $\lqcd$. Now, it is shown how the results can be modified
with running $\alpha_s$. Also, the SCET prediction for
another event shape, the jet mass $\rho$, is given.

It is straightforward to allow $\alpha_s$ to run. Including 1-loop running,
the effect is to modify the
single and double logs in the following way~\cite{Becher:2006mr}:
\begin{equation}
\alp \log^2 \frac{\mu}{\nu} \rightarrow 
- S(\nu,\mu) \equiv -\frac{4\pi  C_F}{\beta_0^2 \alpha_s(\nu)}
\left[ 1 
- \frac{\alpha_s  (\nu)}{\alpha_s (\mu)} 
- \log \frac{\alpha_s (\mu)}{\alpha_s (\nu)} \right]
\end{equation}
\begin{equation}
 \alp \log \frac{\mu}{\nu} \rightarrow 
-  A(\nu, \mu) \equiv - \frac{C_F}{\beta_0}
\log \frac{\alpha_s (\mu)}{\alpha_s (\nu)} .
\end{equation}
For example, the differential thrust distribution for $\tau>0$
with running $\alpha_s$ becomes 
\begin{eqnarray}
  \left[ \frac{1}{\sigma_0} \frac{\rd \sigma}{\rd \tau} \right]_{\mathrm{SCET}}
\!\!\!\!&=&\!  \frac{1}{\tau} \exp\left[
4 S(Q,Q \tau) + 6 A(Q,Q \tau) 
- 8 S(Q\sqrt{\tau},Q \tau) -6 A(Q \sqrt{\tau},Q \tau)\right] 
\tilde{r} (\partial_{\eta}) \frac{e^{- 2 \gamma_E \eta}}{\Gamma [2 \eta]} \nonumber \\
&&  + \frac{2 \alpha_s}{3 \pi} d_{\mathrm{fin}} (\tau),
 \label{scetfinal}
\end{eqnarray}
where
\begin{equation}
  \eta = 4 A (Q \sqrt{\tau},Q \tau) ,
\end{equation}
$d_{\mathrm{fin}}$ is given
in Eq.~\eqref{partondc} and
\begin{equation}
  \tilde{r} (\partial_{\eta}) = 1 + \frac{2}{3\pi}
\left\{
(-8+\frac{7\pi^2}{6} )\alpha_s(Q)
+
(7-\frac{2\pi^2}{3}) \alpha_s(Q \sqrt{\tau}) 
-
\frac{\pi^2}{2} \alpha_s(Q \tau) \right. \nonumber
\end{equation}
\begin{equation}
\left.
+\left[ \frac{1}{2} \alpha_s(Q\sqrt{\tau})
-  \alpha_s(Q \tau)\right] \partial_{\eta}^2 
- \frac{3}{2} \alpha_s(Q\sqrt{\tau})
  \partial_{\eta} \label{r2}
\right\} .
\end{equation}
This is the same function $\tilde{r}$
as in Eq.~\eqref{rtwiddef}, but with the $\alpha_s$ factors evaluated
at the appropriate matching scales.

At this point, one would hope to compare to data. However,
besides the obvious shortcoming of not containing the full
NLL resummation (it does not include effects of the two-loop
cusp anomalous dimension), this parton-level expression
does not include important non-perturbative effects.
Due to the running of $\alpha_s$, the perturbative expression
breaks down when the soft scale is of order $\lqcd$,
that is, when $\tau \sim \lqcd/Q$, as can be seen explicitly in~\eqref{r2}.
 In fact, even
for significantly larger values of thrust power corrections of order
$\lqcd/Q$ become quantitatively important, at least at LEP energies. 
This problem has been approached 
elsewhere using SCET~\cite{Bauer:2002ie,Lee:2006nr}
and with other 
techniques~\cite{Gardi:2001ny,Korchemsky:2000kp,Korchemsky:1999kt}.

Other event shapes can be studied the same way as thrust.
For example, consider the heavy jet mass $\rho$ defined by
\begin{equation}
  \rho \equiv \frac{1}{Q^2}\max (p_L^2, p_R^2).
\end{equation}
In this case, the matching scales are $\mu_h=Q$, $\mu_j=Q \sqrt{\rho}$
and $\mu_s=Q\rho$ and SCET gives for $\rho>0$
\begin{eqnarray}
\left[  \frac{1}{\sigma_0} \frac{\rd \sigma}{\rd \rho}  \right]_\mathrm{SCET}
&=&
\frac{2}{\rho} |1 + c_H |^2
 \exp\left[
4 S(Q,Q \rho) + 6 A(Q,Q \rho) 
- 8 S(Q\sqrt{\rho},Q \rho) - 6 A(Q \sqrt{\rho},Q \rho)
\right] \nonumber\\
&& 
\times \left[ \tilde{k}
  (\partial_{\eta}) \frac{e^{- \gamma_E \eta}}{\Gamma [\eta]} \right] \left[
  \tilde{k} (\partial_{\eta}) \frac{e^{- \gamma_E \eta}}{\Gamma [\eta + 1]}
  \right] + \frac{2 \alpha_s}{3 \pi} d_{\mathrm{fin}} (\rho) ,
\end{eqnarray}
where
\begin{equation}
  \eta = 4 A (Q\sqrt{\rho},Q) .
\end{equation}
This formula  agrees with the jet mass distribution in~\cite{Catani:1992ua}
to leading log and first order in $\alpha_s$.
The same function $d_\mathrm{fin}$ appears for jet mass and
for thrust because to
order $\alpha_s$ in the parton model, $\rho=\tau=\min(\hhs,\hht,\hhu)$.

\section{Conclusions}
We have investigated how to combine resummation with next-to-leading order
matching of event shapes in the original
approach of~\cite{Catani:1992ua} (CTTW) and using SCET. In the CTTW formulation,
real emissions from collinear splitting 
functions are used and various kinematical features
associated with soft emission are combined to derive a differential cross
section. The cross section factorizes into the product of two jet
functions. Resummation is done by solving a differential equation for the jet
functions in terms of the physical scales $p^2$ and $Q^2$ of the
event. In contrast, SCET factorizes the event shape distribution into a
contribution from hard, jet, and soft functions. These functions are matched at
the scales $\mu_h = Q$, $\mu_j = p$ and $\mu_s = \frac{p^2}{Q}$ respectively
and resummation is done through renormalization group evolution. The resummed
thrust distribution in SCET and CTTW have been compared, and found to agree to
next-to-leading order in $\alpha_s$ and for leading-log resummation.

The resummation of thrust in SCET brings to light a number of features of the
process not obvious in CTTW formulation. For example,
the appearance of the seesaw scale $\mu_s = Q (1 - T)$ as the natural matching
scale for the soft function follows from the kinematics of the SCET
factorization theorem. Of course, the existence of this scale has been known
for a long time from QCD, but in the effective field theory
approach this scale just drops out of the factorized expression. Thus, for
more complicated processes, it is reasonable to expect a similar transparency
for the matching scales, which may facilitate subleading log resummation. In
fact, two-loop, three-loop, and some four-loop 
anomalous dimensions for various quantities
are already 
available~\cite{Belitsky:1998tc,Moch:2004pa,Becher:2005pd,Becher:2006qw}, 
and so subleading log resummation appears quite feasible.
The biggest impediment to using these more accurate resummed results in a comparison
to data is that power corrections of order $\lqcd/Q$ have an important
quantitative effect on event shapes. However, these corrections 
should modify only the soft function while higher order resummed expressions
for the the hard and jet functions will remain valid. Thus, the effective
theory should be able to weave together the perturbative and non-perturbative
calculations.

A new result of this paper is the demonstration that inclusive quantities,
such as the total cross section for $e^+ e^-$ can be calculated in a new way
using SCET. Instead of taking the full differential $n$+1 body cross section
and combining with the one-loop $n$-body cross section, SCET combines finite
parts of loops of soft and jet functions with a hard matching calculation and
a finite integral over hard emissions. The soft and jet functions depend on
the infrared regulator, but their convolution, and the hard function, do not.
For $e^+ e^-$ annihilation at NLO, this may not be so impressive, but the
procedure promises to apply to more complicated processes, perhaps even some
for which NLO results are not available. It would also be very interesting to
explore this mechanism at NNLO or to work with hadronic processes where the
singularity structure is more complicated.

\section*{Acknowledgements}
I would like to thank T.~Becher 
and S.~Fleming for many useful discussions. I also benefited
from the hospitality of the Les Houches Ecole de Physique
and discussions with many of its participants.
This work was supported in part by the National Science Foundation under grant
NSF-PHY-0401513 and by the Johns Hopkins Theoretical Interdisciplinary Physics and 
Astronomy Ceneter.

\end{document}